\documentclass[aps,showpacs,superscriptadress,preprint]{revtex4}
\usepackage{graphicx}
\usepackage{epstopdf}

\begin{document}
\title{Stability of the warped black string with nontrivial topology in five-dimensional Anti-de Sitter spacetime}

\author{Shaoyu Yin}
\affiliation{Department of Physics, Fudan University, 200433 Shanghai}

\author{Bin Wang}
\affiliation{Department of Physics, Shanghai Jiao Tong University, 200240 Shanghai}

\author{Robert Mann and Chong Oh Lee}
\affiliation{Department of Physics \& Astronomy, University of Waterloo, Waterloo, Ontario N2L 3G1, Canada}

\author{Chi-Yong Lin}
\affiliation{Department of Physics, National Dong Hwa University, Shoufeng, 974 Hualien}

\author{Ru-Keng Su}
\affiliation{Department of Physics, Fudan University, 200433 Shanghai.}

\begin{abstract}
We investigate the stability of a new warped black string with nontrivial topologies in five-dimensional Anti-de Sitter spacetime. After studying
the linear gravitational perturbation, we find that this black string is unstable when the Kaluza-Klein mass falls in a certain range and the instability exists for all topological spacetimes.
\end{abstract}


\maketitle

\section{Introduction}
String theory makes the radical prediction that spacetime has extra dimensions and gravity propagates in higher dimensions. With the development
of string theory, black holes in higher dimensional space-times have come to play a fundamental role in physics \cite{1} and there is an expectation that
such higher dimensional black holes can be produced either in particle collisions with a mass energy in theTeV range or in the earth's atmosphere due to
the high energy cosmic ray showers \cite{4,5,6}. While the possibility of observing signatures of these kinds of black holes has been discussed before \cite{2,3}, some essential properties of higher dimensional black holes have so far not been fully understood.  Black hole uniqueness theorems,  which are well established for the black hole in four-dimensional spacetimes \cite{Israel:1967,Carter:1971,Hawking:1972,Robinson:1975}, do  not apply to higher dimensional black holes. For example  five-dimensional stationary vacuum black holes are
not unique: there exists the  Myers-Perry solution \cite{1}, which is a generalization of the Kerr solution to arbitrary dimensions, and  rotating black ring solutions with the same angular momenta and mass but different horizon topology \cite{BR}. Even for the $n$-dimensional hyperspherically symmetric asymptotically flat static vacuum Schwarzschild-Tangherlini solution, the uniqueness property fails if one drops the condition of asymptotic flatness but still insists on the same space-time topology. In higher dimensions, there can be more than one solution with non-static or non-flat asymptotic backgrounds, sentencing the higher dimensional uniqueness theorem to death
\cite{13}.

Worse still, the stability of configurations in higher dimensions is also questionable. While four-dimensional black holes are always stable (as
discussed in detail by many authors \cite{Regge:1957,Vishveshwara:1970}), it was first found by Gregory and Laflamme \cite{Gregory:dual} that
black strings and p-branes in asymptotically flat higher dimensional spacetime are unstable under gravitational perturbations. This instability,
often referred to as the Gregory-Laflamme (G-L) instability, is quite general in higher dimensional spacetimes and continues to attract attention
amongst physicists (for a recent comprehensive review, see Ref.\cite{Harmark:2007}) since the detailed nature of this instability, along with its
generalizations to other cases, is still not fully understood. For example an instability was recently reported in charged black holes in higher
dimensional de Sitter spacetime \cite{Konoplya:2009,Cardoso:2009}.

Hence it is of great interest to study stable/unstable configurations in higher dimensional anti-de Sitter (AdS) spacetimes, in part to broaden
the testing ground  for  the proposed AdS/CFT correspondence \cite{Maldacena:1998,Gubser:1998,Witten:1998}, which relates a gravitational theory
on asymptotically AdS spaces in $d + 1$ dimensions to a non-gravitational quantum field theory in $d$ dimensions. The stability of
higher-dimensional AdS configurations has been much less explored relative to their asymptotically flat counterparts. Gregory first studied the
stability of the Randall-Sundrum spacetime and found the G-L instability within a certain range of the $z$ direction mode \cite{Gregory:2000}. An
examination of the stability of the AdS configurations has been extended to the uniform black string solutions including the  exact Schwarzschild
\cite{Hubeny:2002,Hollowood:2007} and Kerr  \cite{Murata:2008} black string solutions to the Type IIB equations of motion in AdS$_5\times S^5$
spacetimes, and approximate black ring solutions in (A)dS spacetime \cite{Caldarelli:2008}. In all of these cases a G-L instability was observed
within a certain parameter range, such as the mode of perturbation or the horizon size or the AdS radius.

Uniform black string solutions are different from the warped AdS black string. For example they have no dependence on the compact extra
dimension.  Although there is no way to  obtain a solution in $d$-dimensional anti de Sitter spacetime by simply adding an extra dimension,
static solutions can be numerically obtained for $d=5$ \cite{Keith} and higher \cite{Radu}  by allowing non-trivial radial dependence of the
metric component associated with this extra dimension.  Stability studies have been carried out for this case \cite{Brihaye:2008} and extended to
nonuniform black string solutions \cite{Delsate:2008}, where only a perturbative dependence on the extra dimension was considered.  Although the
stability of higher dimensional configurations with nonperturbative extra-dimensional dependence, which are exact warped solutions of the
Einstein equations, were first investigated several years ago \cite{Hirayama:2001}, the issue of classical stability of the AdS configurations
has not been addressed extensively. It is generally expected that the AdS solutions can provide a laboratory to test the Gubser-Mitra (GM)
conjecture\cite{16}, which correlates the dynamical and thermodynamical stability for systems with translational symmetry and infinite extent.
A recent attempt to test  this relation for a BTZ black string in 4 dimensions was carried out in \cite{Liu:2008}.

In this paper we further study the stability of  warped AdS black strings in five dimensions. We examine the stability under gravitational tensor
perturbations of both topologically trivial and non-trivial warped AdS black string solutions to the Einstein equations, extending a study of the
scalar perturbation case \cite{Cohlee}.  In higher dimensions the horizon topologies are very rich
\cite{Kol:2002,Kol:2005,Kol:2006,Niarchos:2008,aa,bb,cc}, and their influence on stability can be significant. Furthermore, a
non-asymptotically-flat spacetime can also accommodate richer global topologies. Even an asymptotically AdS four-dimensional black hole can have
horizons with different genus $g$ \cite{Ammin:1996,Mann:1997,Vanzo:1997,mann}.  Studies of quasinormal modes of scalar fields in such spacetimes
indicated that
 the topology of
the four-dimensional AdS black hole does influence stability \cite{CM:1997,Wang:2002}. For the five-dimensional AdS black string and AdS soliton
string with Ricci-flat horizons (genus $g=1$) it has been argued that there is no G-L instability \cite{Chen:2008}. However for the higher
dimensional uniform AdS black string \cite{Radu} with spherical topology the G-L instability was observed for certain values of the AdS radius
whereas the toroidal and hyperbolic spacetimes were found to be stable \cite{Brihaye:2008}. These results show that the topological influence on
the G-L instability in higher dimensional AdS configurations is important. In this paper we investigate the  stability of the warped AdS black
string for different topologies, examining the topological influence on the stability.

Our paper is organized as follows: In the next section we will present the five-dimensional warped AdS topological black string obtained from the
Einstein equation and construct the theoretical framework of the tensor perturbation on its background. We will analyze the analytic solution of
the perturbation equations. In Sec.III, we will numerically solve the perturbation equation and examine the stability of the spacetime.  In the
last section we will present our summary and discussion.

\section{Tensor perturbation of the nontrivial topological black string}

We consider spacetimes that satisfy the Einstein equations with negative cosmological constant $G_{\mu\nu}-\Lambda g_{\mu\nu}=0$. In five
dimensions  the warped AdS black string is an exact solution to these equations
\begin{eqnarray}\label{met}
ds^2&=&-a(z)^2\left(f_k(r)dt^2+\frac{dr^2}{f_k(r)}
+ r^2(d\theta^2+ \Theta_k(\theta)d\phi^2)\right)+dz^2 \\
&=&\frac{1}{L^2\cos^2\psi}\left(f_k(r)dt^2+\frac{dr^2}{f_k(r)}
+ r^2(d\theta^2+ \Theta_k(\theta)d\phi^2) + L^2 d\psi^2 \right)
\end{eqnarray}
where $f_k(r)=r^2/L^2+g-2M/r$, $L$ is the AdS radius and
the warp factor $a(z)=\cosh(z)/L$, with the coordinate transformation
$\cos\psi = \textrm{sech}z$ relating the two metric forms.

Here $k=1,0,-1$ corresponds to three different
topologies:
\begin{equation}
\left\{
  \begin{array}{ll}
    \Theta_1(\theta)=\sin^2\theta, & \hbox{spherical;} \\
    \Theta_0(\theta)=1, & \hbox{toroidal;} \\
    \Theta_{-1}(\theta)=\sinh^2\theta, & \hbox{hyperbolic.}
  \end{array}
\right.
\end{equation}
with genus $g\geq 2$ topological classes obtained upon appropriate identification in the hyperbolic space.

To examine whether this new solution of the Einstein equation is a stable configuration, we consider
linear perturbations on this warped AdS black string, where the gravitational tensor perturbation is taken to have the form \cite{Gregory:dual}
\begin{equation}\label{hpert}
h_{\mu\nu}=e^{\Omega t}u_m(z)\left(
                               \begin{array}{ccccc}
                                 h_{11}(r) & h_{12}(r) &   &   &   \\
                                 h_{12}(r) & h_{22}(r) &   &   &   \\
                                   &   & h_{33}(r) &   &   \\
                                   &   &   & h_{33}(r)\Theta_k(\theta) &   \\
                                   &   &   &   & 0 \\
                               \end{array}
                             \right).
\end{equation}
The time-dependence of the perturbation is chosen with constant frequency $\Omega$. In this form, the perturbation is unstable if the real part
of $\Omega$ is positive, which will be used as a criterion in the stability analysis. This can be compared with the study of the quasinormal
modes, where the time evolution of the perturbation is usually described by $e^{-i\omega t}$.  By comparison we see that our imaginary part
$\Im(\Omega)$ corresponds to the real part of the quasinormal frequency $-\Re(\omega)$ while our real part $\Re(\Omega)$ corresponds to the
imaginary part of the quasinormal frequency $\Im(\omega)$. The dependence of $z$ for each component in the perturbation is uniform and separable.
The gravitational tensor perturbation satisfies the transverse-traceless gauge conditions $h=0$ and $\nabla^\mu h_{\mu\nu}=0$. Applying the
traceless condition, we can express $h_{33}$ in terms of $h_{11}$ and $h_{22}$, and from the transverse constraint we can express $\frac{dh_{12}}{dr}$
and $\frac{dh_{22}}{dr}$ in terms of $h_{11}$, $h_{12}$ and $h_{22}$, thereby reducing the number of independent components in the perturbation tensor.

The linearized Einstein equations yield
\begin{equation}
\nabla^\rho\nabla_\rho h_{\mu\nu}+2R_{\mu\sigma\nu\tau}h^{\sigma\tau}=0.
\end{equation}
Although a total of fifteen equations  must be satisfied by $h_{\mu\nu}$, many of them become trivial once (\ref{hpert}) is used and the
 transverse-traceless
condition is applied. Furthermore, while nontrivial equations are obtained for the $11$, $12$, $22$ and $33$ components,  the equation for the
$33$ component can be recovered from the other three independent equations. From the equations for the $11$, $12$ and $22$ components we can
eliminate $h_{12}$ and $h_{22}$ and finally arrive at a second order ordinary differential equation for $h_{11}$ only. In this equation,  the
equation for $u_m(z)$ separates out with a separation constant $m_K$ that satisfies the equation
\begin{equation}\label{KK}
\cosh^2z\frac{d^2u_m}{dz^2}+(m_K^2-4\cosh^2z)u_m=0.
\end{equation}
This $m_K^2$ represents the influence of the fifth dimension  on the final perturbation equation, and  is usually called the Kaluza-Klein (KK) mass.
We finally arrive at a second order ordinary differential equation for $h_{11}(r)$ with the KK mass $m_K$,
\begin{eqnarray}\label{eqn}
&&\left\{\frac{5m_K^2-10}{4}L^4f_k^3(r)-\left[\frac{9\Omega^2}{4}L^4
+k\frac{3m_K^2-6}{2}L^2-(m_K^2-2)\left(m_K^2-\frac{7}{2}\right)r^2\right]L^2f_k^2(r)\right.\nonumber\\
&&\left.+\left[k\frac{5\Omega^2}{2}L^6+\left(k^2\frac{m_K^2-2}{4}
+\frac{4m_K^2+13}{2}r^2\Omega^2\right)L^4+k\frac{3m_K^2-6}{2}r^2L^2
+\frac{9m_K^2-18}{4}r^4\right]f_k(r)\right.\nonumber\\
&&\left.-\frac{k^2\Omega^2-4r^2\Omega^4}{4}L^6-k\frac{3r^2\Omega^2}{2}L^4
-\frac{9r^4\Omega^2}{4}L^2\right\}r^2L^2f_k^2(r)\frac{d^2h_{11}(r)}{dr^2}\nonumber\\
&+&\left\{\frac{15m_K^2-30}{2}L^4f_k^4(r)-\left[\frac{45\Omega^2}{4}L^4
+k(10m_K^2-20)L^2-3(m_K^2-2)(m_K^2-6)r^2\right]L^2f_k^3(r)\right.\nonumber\\
&&\left.+\left[k\frac{49\Omega^2}{4}L^6+\left(k^2\frac{5m_K^2-10}{2}
+\frac{16m_K^2+79}{4}r^2\Omega^2\right)L^4-k(m_K^2-2)(m_K^2-11)r^2L^2\right.\right.\nonumber\\
&&\left.\left.-3(m_K^2-2)\left(m_K^2-\frac{7}{2}\right)r^4\right]f_k^2(r)
-\left(\frac{3k^2-4r^2\Omega^2}{4}L^4+k\frac{3r^2}{2}L^2
-\frac{9r^4}{4}\right)\Omega^2L^2f_k(r)\right.\nonumber\\
&&\left.-\left(\frac{k+2r\Omega}{2}L^2+\frac{3r^2}{2}\right)
\left(\frac{k-2r\Omega}{2}L^2+\frac{3r^2}{2}\right)
(kL^2+3r^2)\Omega^2\right\}rL^2f_k(r)\frac{dh_{11}(r)}{dr}\nonumber\\
&+&\left\{5(m_K^2-2)L^6f_k^5(r)-\left(k10L^2+\frac{7m_K^2+58}{4}r^2\right)(m_K^2-2)L^4f_k^4(r)\right.\nonumber\\
&&\left.-\left[k\Omega^2L^6-\left(k^2(5m_K^2-10)+\frac{m_K^2+16}{2}
r^2\Omega^2\right)L^4-k\frac{3(m_K^2+10)(m_K^2-2)}{2}r^2L^2\right.\right.\nonumber\\
&&\left.\left.+(m_K^2+4)(m_K^2-2)\left(m_K^2-\frac{7}{2}\right)r^4\right]L^2f_k^3(r)
+\left[\frac{k^28\Omega^2+17r^2\Omega^4}{4}L^8+k(m_K^2-2)r^2\Omega^2L^6\right.\right.\nonumber\\
&&\left.\left.+\left(k^2\frac{(m_K^2-2)^2}{4}-3(m_K^4-m_K^2+4)r^2\Omega^2\right)r^2L^4
+k\frac{3(m_K^2-2)^2}{2}r^4L^2+\frac{9(m_K^2-2)^2}{4}r^6\right]f_k^2(r)\right.\nonumber\\
&&\left.-\left[k\frac{9r^2\Omega^2+2}{2}L^6+3\left(k^2\frac{m_K^2+4}{2}
+\frac{2m_K^2+7}{2}r^2\Omega^2\right)r^2L^4+ 9k (m_K^2+1)r^4L^2\right.\right.\nonumber\\
&&\left.\left.+\frac{27m_K^2r^6}{2}\right]\Omega^2L^2f_k(r)
+\frac{k^2r^2\Omega^4-4r^4\Omega^6}{4}L^8+k\frac{3r^4\Omega^4}{2}L^6
+\frac{9r^6\Omega^4}{4}L^4\right\}h_{11}(r)=0,
\end{eqnarray}
where $k$ here represents the topology, with $k=0$, $1$ and $-1$ indicating the flat, spherical and hyperbolic spaces respectively. The behaviour
of the gravitational perturbation (\ref{hpert})
is determined by this equation. Before solving this perturbation equation numerically, we first
analytically  examine its asymptotic behaviour.

In the asymptotic limit $r\rightarrow\infty$,  eq.(\ref{eqn}) reduces to
\begin{equation}
\frac{d^2h_{11}}{dr^2}-\frac{m_K^2h_{11}}{r^2}=0.
\end{equation}
This equation is independent of $\Omega$ and is universal for all   three topologies. Its solution is
\begin{equation}
h_{11}\rightarrow C_1r^{\frac{1+\sqrt{1+4m_K^2}}{2}}
+C_2r^{\frac{1-\sqrt{1+4m_K^2}}{2}}\quad (r\rightarrow\infty).
\end{equation}
It is clear that a physically reasonable solution should be finite
at infinity. This boundary condition can always be satisfied by
setting $C_1=0$ at $r\rightarrow\infty$ as $m_K^2\geq0$. However near the horizon
eq.(\ref{eqn}) reduces to
\begin{equation}
(r-R_+)^2\frac{d^2h_{11}}{dr^2}+(r-R_+)\frac{dh_{11}}{dr}-\frac{L^4\Omega^2R_+^2}{(3R_+^2+kL^2)^2}h_{11}=0,
\end{equation}
where  $R_+$ is the horizon radius satisfying
$R_+^2/L^2+k-2M/R_+=0$.
The solution is
\begin{equation}
h_{11}\rightarrow C_1(r-R_+)^{\frac{L^2\Omega R_+}{|3R_+^2+kL^2|}}
+C_2(r-R_+)^{-\frac{L^2\Omega R_+}{|3R_+^2+kL^2|}}\quad
(r\rightarrow R_+).
\end{equation}
We notice that the two linearly independent parts are symmetric under sign-reversal of  $\Omega$. For  unstable perturbations with
$\Re(\Omega)>0$, we set $C_2=0$ to ensure finiteness near the horizon.

 Putting both of these boundary conditions together, it would appear that the asymptotic behaviour permits the existence of an unstable
perturbation. However to obtain a definitive answer concerning the stability of the AdS black string (\ref{met}), we have to match these two
asymptotic solutions in the intermediate range. This entails solving the original differential equation eq. (\ref{eqn}) for which we must rely on
numerical calculations.

\section{Numerical solution}

We now turn to the task on solving the gravitational perturbation equation (\ref{eqn}) on the background of the AdS black string numerically. We will use the
numerical method first proposed by Horowitz and Hubeny \cite{Horowitz:2000} and later extensively applied in \cite{he:45} to solve the
perturbation equation. Noticing that the equation is unchanged under the scale transformation $L\rightarrow\alpha L$, $M\rightarrow\alpha M$,
$\Omega\rightarrow\alpha/\Omega$ and $r\rightarrow\alpha r$, we can choose $L=1$ without loss of generality.
Substituting $r=1/x$, one gets a second order differential equation for $h_{11}(x)$ in the range $0<x<X$, where $X=1/R_+$. The perturbation
equation we need to solve can be changed into
\begin{equation}\label{h11eq}
s(x)\frac{d^2h_{11}}{dx^2}+\frac{t(x)}{x-X}\frac{dh_{11}}{dx}+\frac{u(x)h_{11}}{(x-X)^2}=0,
\end{equation}
where three coefficients can be expanded around $x=X$ as,
\begin{equation}
s(x)=\sum_{n=0}^\infty s_n(x-X)^{n},\quad t(x)=\sum_{n=0}^\infty t_n(x-X)^{n},\quad u(x)=\sum_{n=0}^\infty u_n(x-X)^{n}.
\end{equation}
Equation (\ref{h11eq}) can be solved by series expansion about $x=X$,
\begin{equation}
h_{11}(x)=\sum_{n=0}^\infty a_n(x-X)^{n+\rho},
\end{equation}
where $\rho$ is the lowest order of $h_{11}$ in the expansion.
Following standard mathematical methods, the recursion relation for $a_n$ can be obtained in the form
\begin{equation}
a_n=-\frac{\sum_{k=0}^{n-1}[(k+\rho)(k+\rho-1)s_{n-k}+(k+\rho)t_{n-k}+u_{n-k}]a_k} {(n+\rho)(n+\rho-1)s_0+(n+\rho)t_0+u_0},
\end{equation}
where $\rho$ is determined by the equation $s_0\rho(\rho-1)+t_0\rho+u_0=0$ to be $\rho=\pm\frac{X\Omega}{3+gX^2}$. Notice that one should
choose the solution of $\rho$ positively proportional to $\Omega$, which corresponds to the ingoing wave at the horizon \cite{Horowitz:2000}.

The boundary condition at infinity is now $h_{11}(0)=\sum_{n=0}^\infty a_n(-X)^{n+\rho}=0$. Since the summation depends on the value of $\Omega$
and $m$, the problem is transformed into finding the zeros of $\sum_{n=0}^\infty a_n(-X)^{n+\rho}$ as function of $\Omega$ for a fixed $m$. If
the value of $\Omega$ lies in the right half of the complex plane, according to the time dependent part of our ansatz $e^{\Omega t}$, the black
string is unstable under such perturbation. Of course, the summation of infinite terms is impossible in a real calculation. We instead stop the
summation at  finite $n$, taken to be sufficiently large  so as to ensure that the solution will remains stable even as the summation goes beyond
this $n$. The convergence of the result becomes slower for small $m_K$ which needs a larger value of $n$ in the partial sum to reduce the
relative error in the computation.  Since this takes a lot of computer time  we employ the trial and error method proposed in \cite{lin:11} to
truncate the sum at some large $n$. However whatever  large value of $n$ we adopt, an initial tiny error may grow through recursion relations. We
therefore improve the precision of all input data with the help of a built-in function of mathematica. In our calculation we take the truncation
at $n=350$ for most cases, which ensures the precision of the numerical calculation for $\Omega$ is
not worse than $10^{-10}$ when the KK mass is
around or bigger than unity. When the KK mass is very small, the accuracy for the calculation of $\Omega$ will be reduced, however with our
selected truncation number we can still keep the accuracy to be higher than $10^{-3}$ when the KK mass is very small.

Now we describe  general features of our numerical results. We find that in most situations $\Omega$ has a negative real part, and that the larger this part is, the larger the imaginary part will be. In our ansatz $e^{\Omega t}$, the imaginary part of $\Omega$ represents the
energy of the perturbation. Thus in most cases the gravitational perturbation has stable decay modes and the perturbation with higher energy
decays faster. The only exception is  the case with $k=-1$ and $M=0$, where most solutions of $\Omega$ are purely real negative without imaginary
part, which means that the gravitational perturbation decays without oscillation. Similar behavior has also been reported for the hyperbolic
four-dimensional AdS black hole under scalar perturbations \cite{Wang:2002}.

Besides these observed decay modes, we also obtain purely positive real values for $\Omega$ in some regions of parameter space, which represents
the purely growing mode of the gravitational perturbation.  These purely real $\Omega$s are the minimum value upon comparison with other modes
$\Omega=\sqrt{\Omega_R^2+\Omega_I^2}$ \cite{Maggiore:2008}, which indicates that they are always the ground state perturbative modes. Their
existence signals an instability.   The growing modes of the perturbation in the ground states appear for a certain range of the KK mass for
different topologies and sizes of the warped AdS black string. In table I we list the characteristic range of the KK mass for different
topological AdS black strings when their radii are bigger than, equal to or smaller than the AdS radius, respectively. It is clear that the
growing modes exist for all topologies. This observation supports our approximate analysis in the last section.

\begin{table}
\caption{The range of KK mass yielding instability for different black string topologies and sizes}
\begin{tabular}[t]{ccccccccc}\hline
\centering
Topology & $M$ & $R_+$ & $\rho$ & range of the KK mass \\
\hline\hline
$k=1$ & $\frac{5}{16}$ & $\frac{1}{2}$ & $\frac{2\Omega}{7}$ & $2<m_K^2<2.7$ \\
\hline
$k=1$ & $1$ & $1$ & $\frac{\Omega}{4}$ & $0.6<m_K^2<2$ \\
\hline
$k=1$ & $5$ & $2$ & $\frac{2\Omega}{13}$ & $0.135<m_K^2<2$ \\
\hline\hline
$k=0$ & $\frac{1}{16}$ & $\frac{1}{2}$ & $\frac{2\Omega}{3}$ & $m_K^2<2$ \\
\hline
$k=0$ & $\frac{1}{2}$ & $1$ & $\frac{\Omega}{3}$ & $m_K^2<2$ \\
\hline
$k=0$ & $4$ & $2$ & $\frac{\Omega}{6}$ & $m_K^2<2$ \\
\hline\hline
$k=-1$ & $-\frac{3}{16}$ & $\frac{1}{2}$ & $2\Omega$ & no boundary found \\
\hline
$k=-1$ & $0$ & $1$ & $\frac{\Omega}{2}$ & $0<m_K^2<2$ \\
\hline
$k=-1$ & $3$ & $2$ & $\frac{2\Omega}{11}$ & $m_K^2<2$ \\
\hline
\end{tabular}\label{table1}
\end{table}

For the spherical spacetime with $k=1$, we see that as the horizon size increases, the range of KK mass accommodating unstable modes also
increases. This property is also exhibited in Fig.1, which is consistent with the left panel of Fig.1 in \cite{Hirayama:2001} where only the
stability of the spherical AdS black string is studied. For the $k=0$ case  we find that the ranges of $m^2_K$ that allow instability are
independent of horizon radius, as shown in Fig.2. When the spacetime is of higher genus (hyperbolic with $k=-1$) the warped AdS black string
contains richer physics. It can allow the black string to contain negative mass, which is also seen for the hyperbolic AdS black hole case
\cite{Mann:neg}, with horizon smaller than the AdS radius. In this situation, the gravitational perturbation is always unstable regardless of the
value of the KK mass. This result agrees with the finding in the four-dimensional AdS spacetime obtained by Hawking and Page \cite{Hawking:1983}.

For the hyperbolic topology, when $M=0$, we have the AdS black hole of zero mass and the horizon $R_+=1$ is just the AdS radius of unit $L$. If
the KK mass vanishes (ie. $m_K=0$, which means that the extra dimensions have no influence) the space itself is stable, as   shown in Table 1.
Interestingly, when the effects of extra dimensions are turned on  the gravitational perturbation becomes unstable. This can be seen even for very
small values of $m_K^2$, and this instability lasts until $m_K^2\sim 2$. For the positive mass AdS black string, the range of the KK mass  allowing instability is shown in table I. In Fig.3 we show the unstable range of $m_K^2$ for zero mass and positive mass hyperbolic AdS black
strings.

In all cases $m_K^2=2$ is special; at this point the positive real part of $\Omega$ vanishes. The peculiar value $m_K^2=2$ can be understood by looking at
eq.(\ref{KK}) where the general solution to the differential equation of $u_m(z)$ has the very simple form
$C_1(\cosh(2z)+1)+C_2(\cosh(2z)+2)\sinh(2z)/(\cosh(2z)+1)$ rather than generally the very complicated hypergeometric functions, and many terms in
eq.(\ref{eqn}) vanish when $m_K^2=2$. The complexity of eq.(\ref{eqn}) puts practical limitations on exploring the full parameter space of
continuously varying $M$ or $R_+$ with regard to stability. However the behaviours presented here reflect the general properties of the stability of this
warped AdS black string with different topologies.

\begin{figure}
\resizebox{0.7\linewidth}{!}{\includegraphics*{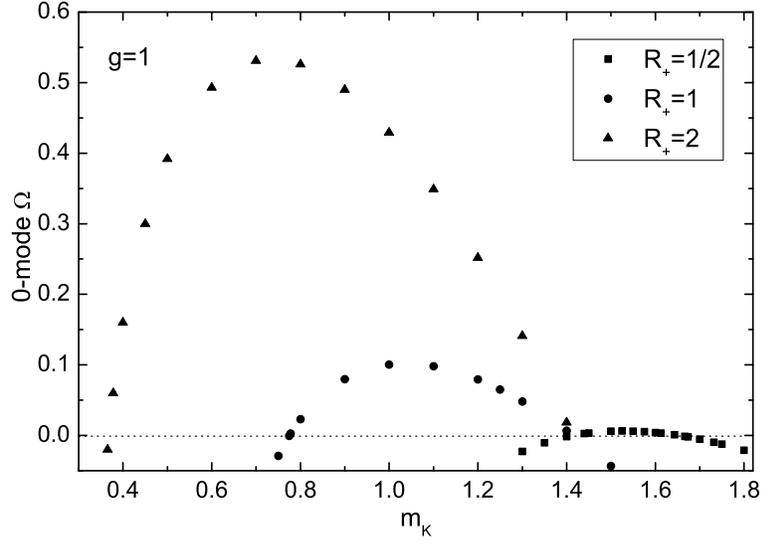}} \caption{The range of KK mass  allowing instability for a spherical AdS black
string.}\label{g1}
\end{figure}
\begin{figure}
\resizebox{0.7\linewidth}{!}{\includegraphics*{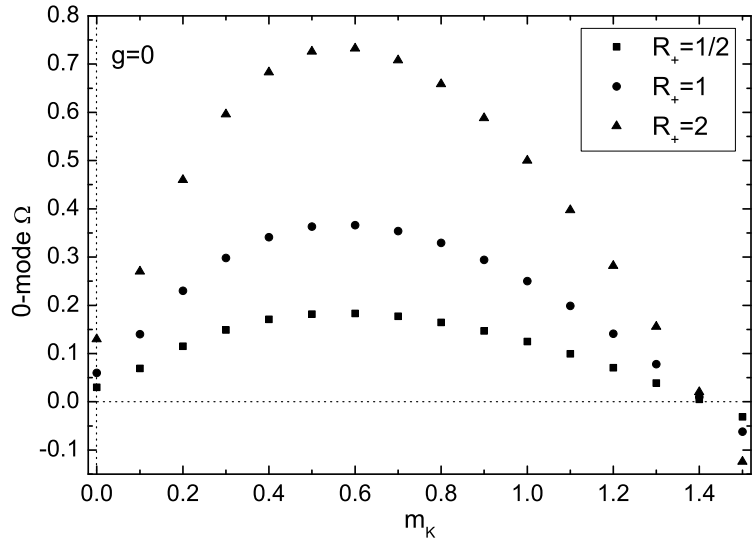}} \caption{The range of KK mass allowing instability for a flat AdS black string.}\label{g0}
\end{figure}
\begin{figure}
\resizebox{0.7\linewidth}{!}{\includegraphics*{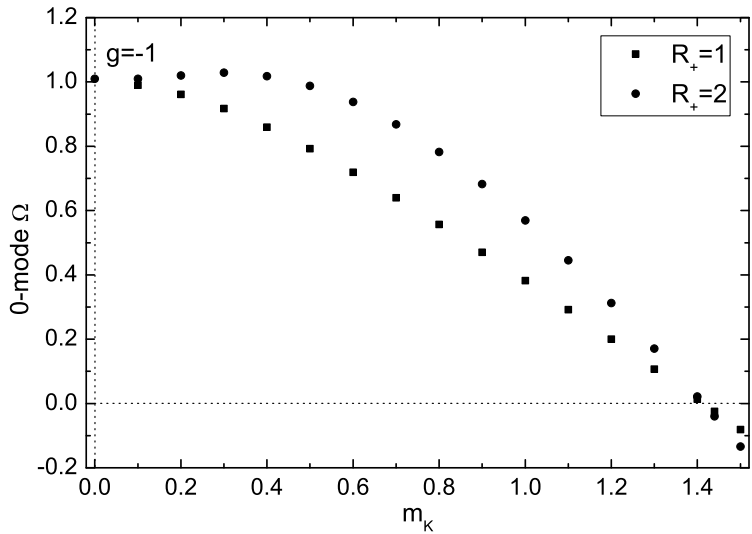}} \caption{The range of KK mass allowing instability for a hyperbolic AdS black string
with zero and positive mass.  } \label{g-1}
\end{figure}

\section{Summary and discussion}

We have examined the stability of a new warped black string in five-dimensional AdS spacetime with nontrivial topologies obtained by solving
Einstein equations. From the numerical calculation, we have shown that the black string can be unstable for all topologies provided that the KK
mass falls within a certain range. The influence of the topology and extra dimension on the instability has been clearly exhibited.

The stability of the AdS black string solution is an interesting topic. It provides a new laboratory to test the Gubser-Mitra conjecture
\cite{16}, that correlates the dynamical and thermodynamical stability for systems with translational symmetry and infinite extent. This
conjecture has passed a large number of tests \cite{Liu:2008, Hirayama:2001}, but it is also known to fail in certain cases \cite{18}.

Our research here has been completely dynamical. It will be interesting to see how the instability we observed here can be related to the
thermodynamic instability. Noticing that usually the thermodynamic discussion cannot involve the KK mass. Understanding the relation between the
dynamic instability and the thermodynamic instability of black strings remains an interesting topic for further study that we hope will be
carried out in the future.

\section*{Acknowledgements}

This work is supported in part by NNSF of China and by the Natural Sciences and Engineering Research Council of Canada. S. Yin is partially
supported by the graduate renovation foundation of Fudan University.

\end{document}